\documentclass[11pt]{article}
\usepackage[margin=1in]{geometry}
\usepackage[T1]{fontenc}
\usepackage[utf8]{inputenc}
\usepackage{mathptmx}
\usepackage[expansion=false]{microtype}
\usepackage{setspace}
\usepackage{amsmath}
\usepackage{pdflscape}
\usepackage{graphicx}
\graphicspath{{plots/}}
\usepackage{booktabs}
\usepackage{paracol}
\usepackage{tabularx}
\usepackage{array}
\usepackage{caption}
\usepackage{float}
\usepackage{lineno}
\usepackage[numbers,sort&compress,super]{natbib}
\usepackage[hidelinks]{hyperref}
\usepackage{url}
\usepackage{xcolor}
\hypersetup{
	pdftitle={More of the same? Are scientific papers losing originality?},
	pdfauthor={Ilan Doron-Arad and Elchanan Mossel},
}
\title{More of the same? Are scientific papers losing originality?}
\author{Ilan Doron-Arad$^{1,*}$ and Elchanan Mossel$^{1}$\\[0.4em]
	\small $^1$Department of Mathematics, Massachusetts Institute of Technology, Cambridge, MA, USA\\
	\small $^*$Correspondence: \href{mailto:ilanda@mit.edu}{ilanda@mit.edu}} 
	\date{}
	
\begin{document}
\maketitle

\begin{abstract}
	Scientific output is growing rapidly,\cite{bornmann2021} but it is unclear whether the expanding literature remains original or is increasingly repeating itself. Originality has many dimensions. 
    One dimension can now be measured directly: how textually distinct a paper is from the work that came before it. Using semantic language models, we screened more than 26,000 full-text papers across eight fields in the CORE database (2015--2025) for passages resembling recent literature. We compared every paper against a fixed-size sample of earlier work, so increases in resemblance are not a by-product of literature growth. We find that the share of a paper's text resembling recent work was stable before 2021, then rose sharply, more than tripling by 2025. The increase came not from a few papers borrowing more heavily but from resemblance spreading across the literature: more papers now contain passages that echo recent work, and almost none of this growth is explained by cited work. We also found an analogous increase in passages expressing the same scientific idea.
    The pattern held across robustness checks and replicated in a second, independent database. Along this one measurable axis of originality, scientific writing has become less distinct from the work preceding it.
\end{abstract}

Science advances by adding something new. Yet the scientific literature is expanding rapidly: publication output grows by roughly 4\% annually, doubling about every 17 years.\cite{bornmann2021} It is not obvious that originality is keeping pace. A larger literature could mean a broader frontier of ideas; equally, it could mean more of the same: more papers that increasingly echo work already published. Whether the growing record remains original or more repetitive is a basic question about the performance of the scientific community.

It is difficult to measure originality, which has many dimensions and no single agreed measure. Yet, one dimension can now be measured at scale. Modern language models can represent the meaning of a passage as a numerical vector and detect when two passages express closely related content, even when they share few of the same words.\cite{sbert,mpnet} This makes it possible to measure the \emph{textual distinctiveness} of a paper: how much of its text is semantically similar to recent literature. Textual distinctiveness is not the same as conceptual novelty: in principle, a genuinely new finding can be written by combining passages from existing literature. Conversely, unusual phrasing need not signal a new idea. Textual distinctiveness is, however, one concrete and measurable facet of originality, and a decline in it may signal a broader decline in originality.

Prior work suggests that science is concentrating around established work. For example, electronic access has narrowed attention to a smaller set of established papers, large fields  can become focused on familiar research directions, and citation patterns suggest that papers are becoming less disruptive over time.\cite{evans2008,chu2021,park2023} This latter interpretation has been questioned because the disruption index can be sensitive to citation inflation.\cite{petersen2024} More broadly, previous work has studied novelty through citation linkage analysis, semantic representations, and the introduction and spread of new scientific concepts.\cite{uzzi2013,funk2017,wang2017,shibayama2021,yin2023,cheng2023,arts2025,peng2025,liu2026} Recent semantic approaches to scientific novelty have generally operated at the paper level, often using titles and abstracts, rather than identifying matched passages within full scientific texts.\cite{shibayama2021,yin2023,hou2022,arts2025,culbert2025,peng2025,safari2026,wang2026semantic} Additionally, other works have studied direct text reuse and, more recently, the growing effects of LLMs on scientific writing.\cite{citron2015,gienapp2023,maclaughlin2021,liang2025,kobak2025,siler2026,filimonovic2026,kousha2026,sourati2026,qian2026} Yet, to our knowledge, no study has determined whether full scientific papers are becoming increasingly semantically similar to recent work over time while holding fixed the number of prior papers used for comparison. This distinction is crucial given the massive growth in literature volume.\cite{bornmann2021}

We address this question using more than 26,000 full-text papers from eight fields in the CORE database,\cite{core} the world’s largest open-access research collection. For each paper we measure how much of its text semantically resembles the recent literature, comparing every paper in 2015-2025 against a fixed size sample of work from the preceding five years. 
Keeping this sample fixed is crucial. With a growing comparison pool, similarity could rise because each passage has more chances to find a similar passage in prior work. By holding the pool size constant, an increase is less likely to reflect literature growth and is more likely to reflect a real change in how papers are written.

\begin{figure*}[t]
    \centering

    \begin{minipage}[t]{0.31\textwidth}
        \textbf{a}\par\vspace{1mm}
        \centering
        \includegraphics[width=\linewidth]{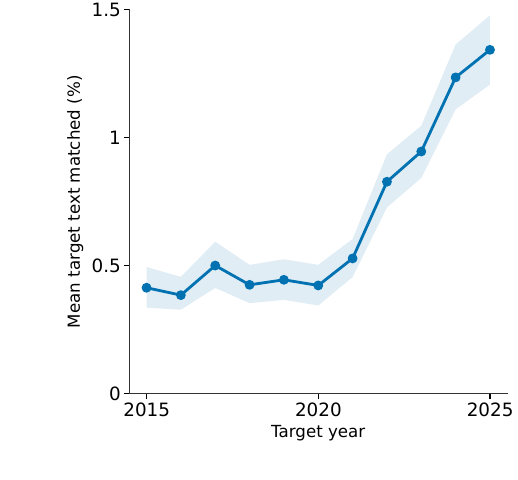}
    \end{minipage}
    \hfill
    \begin{minipage}[t]{0.31\textwidth}
        \textbf{b}\par\vspace{1mm}
        \centering
        \includegraphics[width=\linewidth]{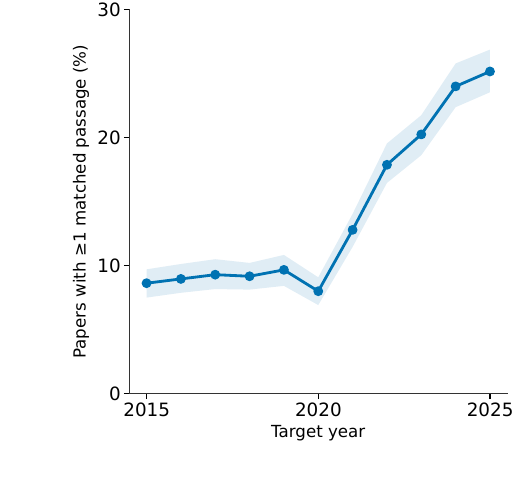}
    \end{minipage}
    \hfill
    \begin{minipage}[t]{0.31\textwidth}
        \textbf{c}\par\vspace{1mm}
        \centering
        \includegraphics[width=\linewidth]{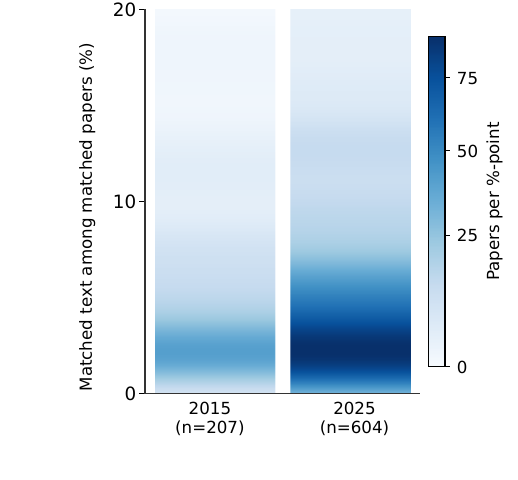}
    \end{minipage}

    \vspace{-2mm}

    \begin{minipage}[t]{0.85\textwidth}
        \textbf{d}\par\vspace{1mm}
        \centering
        \includegraphics[width=\linewidth]{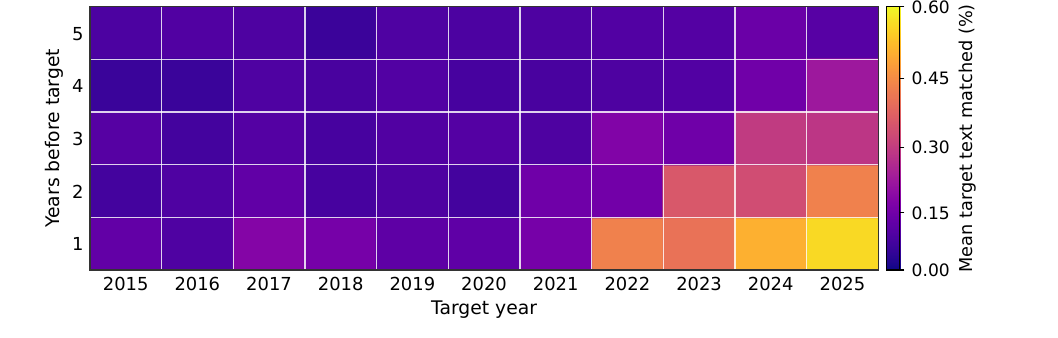}
    \end{minipage}

 \caption{\textbf{Semantic similarity to prior literature increased and spread to more papers.}
\textbf{a}, Annual equal-weight mean percentage of target text matched to prior sources across the eight fields; shaded bands indicate 95\% bootstrap confidence intervals over target papers.
\textbf{b}, Annual equal-weight mean share of papers containing at least one matched passage. 
\textbf{c}, Distribution of target text matched among papers with at least one match in 2015 and 2025; darker shading indicates a greater density of matched papers.
\textbf{d}, Each cell corresponds to one target year and one difference of 1--5 years between the target and source publication years. Its value is the equal-weight mean percentage of target text matched across the eight queries using only sources published at that exact year difference. For example, the cell for 2025 and 2 years uses only sources published in 2023.
}
    \label{fig:main}
\end{figure*}

\begin{figure*}[t]
\centering
\noindent
\begin{minipage}[t]{0.4\textwidth}
    \vspace{0pt}
    \raggedright\textbf{a}\par
    \vspace{1mm}
    \centering
    \includegraphics[width=\linewidth]{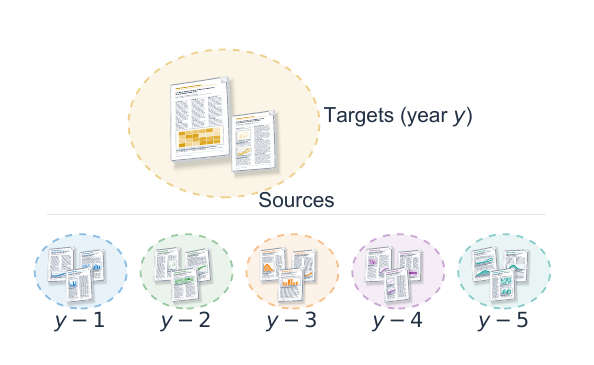}
\end{minipage}%
\hspace{0.01\textwidth}%
\begin{minipage}[t]{0.29\textwidth}
    \vspace{0pt}
    \raggedright\textbf{b}\par
    \vspace{1mm}
    \centering
    \includegraphics[width=\linewidth]{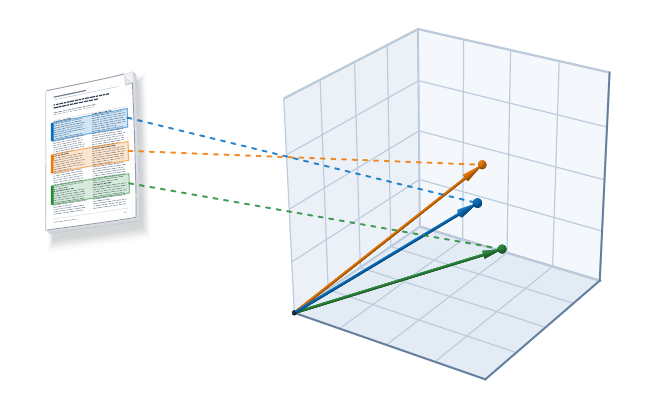}
\end{minipage}%
\hspace{0.01\textwidth}%
\begin{minipage}[t]{0.29\textwidth}
    \vspace{0pt}
    \raggedright\textbf{c}\par
    \vspace{1mm}
    \centering
    \includegraphics[width=\linewidth]{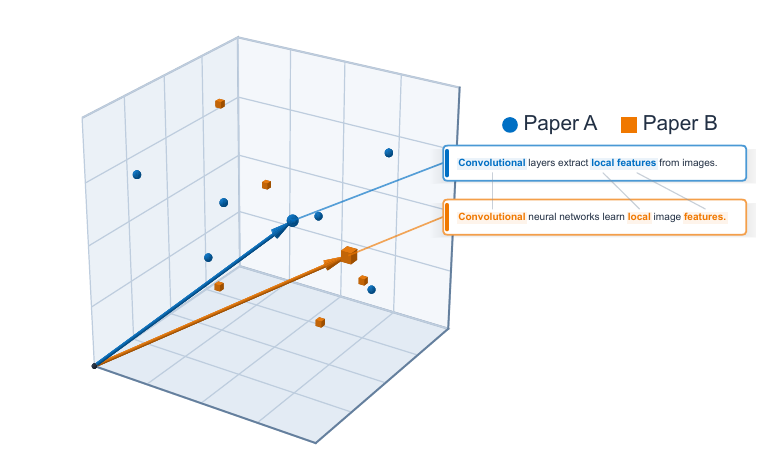}
\end{minipage}

\caption{\textbf{Illustration of the semantic similarity pipeline.}
\textbf{a}, Target papers from year $y$ are compared against a fixed number of source papers from each of the preceding five years.
\textbf{b}, Passages are embedded as vectors in a shared semantic space, shown schematically in three dimensions (the analysis uses 768 dimensions).
\textbf{c}, Similarity between passages is assessed by the angle between their embedding vectors and additional lexical constraints, illustrated for a matched pair of passages.}
\label{fig:pipeline}
\end{figure*}

We find that the share of a paper's text resembling recent work was approximately stable in the years preceding 2021, then rose sharply, more than tripling by 2025 (Fig.~\ref{fig:main}; the pipeline is illustrated in Fig.~\ref{fig:pipeline}). The increase came not from a few papers resembling prior work more heavily but from resemblance spreading across the literature: far more papers now contain passages that echo recent work, and almost none of this growth is explained by cited work. We also find an analogous increase in passages expressing the same scientific idea. The pattern survived extensive robustness checks and replicated in a second, independent database (Europe PMC).\cite{europepmc} Hence, along one measurable axis of originality, scientific writing has become less textually distinct from recent work.

\subsection*{Similarity spread to more papers}

For each field and year, we compared 300 target papers with 3,000 papers from the preceding five years. The eight fields were defined by CORE queries: \emph{astrophysics, bioinformatics, chemistry, climate change, machine learning, medicine, psychology}, and \emph{quantum computing}.
Our primary outcome is semantic similarity coverage: the fraction of target text matched to at least one source passage.

Equal-field mean similarity coverage was approximately stable at 0.432\% in 2015--2020 and then increased from 2021 onward, reaching 1.343\% in 2025, 3.11 times the 2015--2020 mean (95\% bootstrap CI, 2.73--3.52; Fig.~\ref{fig:main}a). Under the same fixed screen, the share of papers containing a similar passage also increased sharply, reaching 25.2\% in 2025 (Fig.~\ref{fig:main}b). 
In all eight fields, coverage in 2025 exceeded the 2015--2020 mean, although the magnitude and year-to-year trajectory varied substantially across fields (the eight trajectories are shown in Extended data Fig.~\ref{fig:main_trends} and~\ref{fig:field_prevalence}). Among matched papers, coverage changed comparatively little, showing that the increase came mainly from similarity spreading to more papers (Fig.~\ref{fig:main}c). An exact one-sided permutation test supported this temporal pattern, identifying 2021 as the best-fitting onset of a persistent increase ($P$ value \(1.80\times10^{-5}\); see Methods).

\begin{figure}[H]
    \centering

    \begin{minipage}[t]{0.31\linewidth}
        \textbf{a}\par\vspace{1mm}
        \centering
        \includegraphics[width=\linewidth]{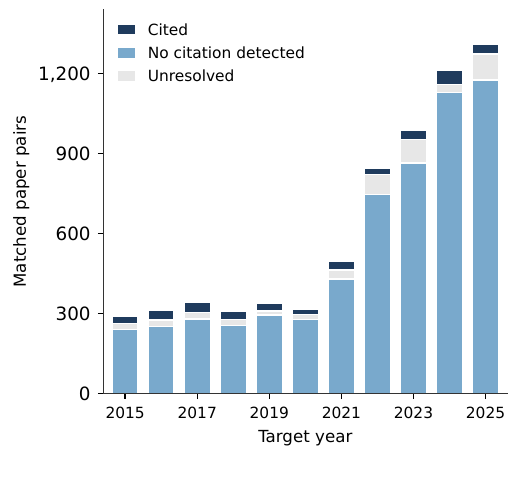}
    \end{minipage}
    \hspace{0.02\linewidth}
    \begin{minipage}[t]{0.31\linewidth}
        \textbf{b}\par\vspace{1mm}
        \centering
        \includegraphics[width=\linewidth]{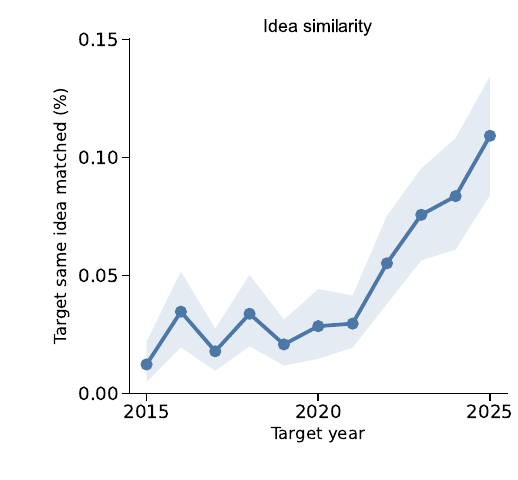}
    \end{minipage}
    \hspace{0.02\linewidth}
    \begin{minipage}[t]{0.31\linewidth}
        \textbf{c}\par\vspace{1mm}
        \centering
        \includegraphics[width=\linewidth]{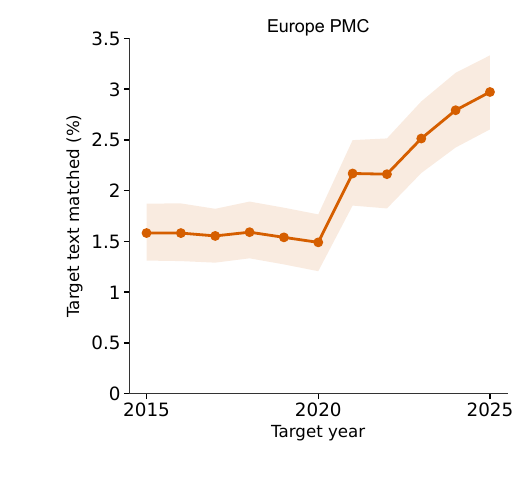}
    \end{minipage}

    \caption{\textbf{Most matches are not cited; idea similarity increases too; the trend replicated in Europe PMC database.}
    \textbf{a}, Annual number of matches in CORE by citation status.
    \textbf{b}, Annual equal-field mean percentage of target text covered by matched passages classified as expressing the same scientific idea.
    \textbf{c}, Annual equal-weight mean semantic similarity target text matched (coverage) across three Europe PMC queries. Shaded bands in b and c indicate 95\% bootstrap confidence intervals over target papers (1,000 bootstrap resamples).}
    \label{fig:interpretation_replication}
\end{figure}

\paragraph*{Most matches are not cited.}
Citation linkage was detected for only 5.3\% of unique matched paper pairs. Despite a more than fourfold increase in the total number of matched pairs, from 290 in 2015 to 1,310 in 2025, the number with a detected citation link increased from 27 to 36 (Fig.~\ref{fig:interpretation_replication}a). Thus, nearly all of the growth occurred among pairs without detected citation linkage. A human review of 176 matched paper pairs, sampled two pairs from each field and year, yielded an estimated citation rate of 6.7\% (95\% stratified bootstrap CI, 2.8–10.6\%), excluding two unresolved pairs, consistent with the automated estimate of 5.3\%.

\paragraph{Idea similarity increases too.}
So far, we have characterized literature convergence in terms of semantic similarity. We next asked whether this trend extends to the level of scientific \emph{ideas}. We performed a post hoc analysis of the detected matches, using a conservative LLM classification designed to favor precision over recall (see Methods). The analysis identified 614 pairs as expressing essentially the same scientific proposition. There were 155 such pairs in 2015--2020 and 459 in 2021--2025, corresponding to a 3.55-fold increase in their annual number. In 2025 alone, 139 same idea pairs were detected, 5.38 times the annual average during the 2015--2020 baseline (Fig.~\ref{fig:interpretation_replication}b). This analysis captures similarity at the level of individual scientific propositions expressed in passages, rather than the novelty of a paper's overall contribution.

\paragraph{The pattern replicates on a second database}

We repeated the experiment on Europe PMC, an independent full-text scientific database, using three queries: \emph{machine learning, climate change,} and \emph{psychology}.\cite{europepmc} The replication used the same years, design and detector. 
Equal-query mean semantic similarity coverage was roughly stable
at 1.556\% in 2015--2020 and subsequently increased, reaching 2.972\%
in 2025, 1.91 times the 2015--2020 mean
(95\% bootstrap CI, 1.65--2.19;
Fig.~\ref{fig:interpretation_replication}c).
 Total similarity increased in all three queries.
Among matched pairs with resolved citation status, only 8.9\% had detected citation, while citation status remained unresolved for 2.6\% of pairs. The CORE and Europe PMC samples had very little overlap in target or source papers, indicating that the replication cannot be explained by analyzing the same papers in both databases (see Methods).

\paragraph{Validation and negative controls.}

We used three complementary diagnostics to address whether the trend is an artifact of the experiment. 
The semantic screens in the first two controls used a broader matching rule (see Methods). 
First, we compared the detector against a direct wording overlap, which remained substantially below the semantic similarity, indicating that the signal is not simply a rise in copied or repeated phrases and terms (Extended Data Fig.~\ref{ed:lexical_semantic}a).  
Second, as a negative control, we compared papers with unrelated older papers from a different topic. Similarity was near zero, whereas comparisons with older papers on the same topic produced substantially higher similarity (Extended Data Fig.~\ref{ed:topic_null}b).
Third, we tested the detector on a synthetic benchmark in which altered versions of the same passage were treated as matches and unrelated passages as non matches. At the fixed threshold, it achieved 99.6\% precision, 81.1\% recall and 89.4\% F1, and correctly classified additional examples designed to fall just above or below the threshold. Thus, in this benchmark, the detector produced very few false positives but missed some true matches (Extended Data Fig.~\ref{ed:calibration}a,b).

\paragraph{Robustness of the trend.}
The increase persisted after accounting for temporal changes in the composition of the CORE sample, including differences in paper length, retrieval position, journal and publisher mix, and title similarity (Extended Data Fig.~\ref{ed:composition_controls}a,b; see Methods).
The pattern also remained across different source pool sizes and semantic similarity thresholds of the screen (Extended Data Figs.~\ref{ed:pool_sweep}c and~\ref{ed:threshold_sensitivity}). The increase was also robust to the maximum passage chunk length used in the analysis (Extended Data Fig.~\ref{ed:chunk_robustness}d). An additional sensitivity analysis controlling for the number of
source and target passages yielded a 2.92-fold increase
in expected coverage from 2015--2020 to 2025
(0.377\% to 1.101\%; 95\% bootstrap CI, 2.56--3.32).

For a common robustness contrast, we compared 2015--2020 with 2021--2025. Mean similarity coverage increased from 0.432\% to 0.976\%, a 2.26-fold increase (95\% bootstrap CI, 2.08--2.48). The increase remained positive across CORE retrieval order deciles, after omitting each field or year in turn, and separately among the higher and lower ranked halves of the sampled CORE results (Extended Data Fig.~\ref{ed:rank_period_robustness}). Restricting matches to sources published at each exact difference between the target and source years, from one to five years, also preserved the increase (Fig.~\ref{fig:main}d). The trend further persisted when sources were restricted to papers never analyzed as targets, after excluding technical methods passages, and after removing frequently matched source papers or passages. We performed an additional sensitivity analysis on \emph{machine learning}, checking the trend with and without a filter that removes passages identified as generic ML content. In this analysis, the ratio of 2025 coverage to the 2015 to 2020 baseline was 2.78 times with the filter and 3.41 times without it, indicating that the filter reduced rather than explained the increase.

To test whether the trend reflected increased crowding of the semantic embedding space, we examined inter-paper passage similarities in \emph{bioinformatics}. The frequency of randomly sampled inter paper passage pairs exceeding cosine 0.84 remained very low and roughly stable across years, with no sustained increase paralleling the rise in similarity coverage. Thus, the observed trend is unlikely to result simply from later passages becoming more densely packed near the matching threshold (Extended Data Fig.~\ref{ed:crowding}).

As a complementary analysis, we also examined arXiv papers from the intersection of \emph{cs.CL}, \emph{cs.LG} and \emph{cs.AI}. Unlike the CORE analysis, each target sample was compared with all available papers from the preceding five years, so the source pool was allowed to grow over time. Mean semantic similarity coverage increased from an average of approximately 0.12\% in 2015--2020 to approximately 0.34\% in 2025, a roughly 2.8-fold increase. Thus, the same qualitative trend was also observed when similarity was measured against the expanding recent literature rather than a fixed-size source pool. Because the available arXiv source pool grew substantially over this period, we treat this analysis as a qualitative robustness check rather than as an estimate directly comparable to the main fixed-pool analysis.

\subsection*{More of the same}

Our results reveal a broad change in the sampled scientific record: as the literature expands, a growing share of papers contains passages semantically similar to those in recent work. 
In the primary analysis, coverage in 2025 exceeded the 2015--2020
mean in all eight studied CORE queries. The aggregate increase
persisted across the robustness checks, extended to passages
classified as expressing the same scientific idea, and was reproduced
in an independent analysis with the Europe PMC database.
The increase was driven mainly by more papers showing similarity, not by stronger similarity among papers having at least one match.
Moreover, the increase occurred almost entirely among pairs without detected citation linkage, indicating that reuse that includes a citation does not explain the trend. Nevertheless, semantic similarity alone does not imply misconduct.

This rise is not simply a consequence of the literature growing. Each paper was compared against a reference sample of fixed size, so the pool of work available to match against did not expand over time. The increase is therefore unlikely to reflect the space of published work merely being packed more densely, and instead points to a change in the papers themselves. 
The arXiv analysis provides a complementary view: similarity also increased when target papers were compared with the full, expanding pool of recent work. Unlike the fixed source CORE pool size, however, this analysis does not separate changes in the papers from the increasing number of opportunities to find a match.

The largest increase coincides with the rapid adoption of LLM-assisted scientific writing.\cite{liang2025,kobak2025,siler2026} 
LLM use is a plausible contributor because it may spread conventional scientific phrasing at scale, especially given that the semantic similarity spread to more papers. However, our study does not establish the main cause of this trend. A broader convergence toward shared disciplinary language could also explain in part the observed pattern. 
More broadly, we read this trend as one visible sign of a wider convergence in how science is written and framed. Namely, several forces plausibly push in the same direction: electronic access narrows attention to a shared set of papers, disciplinary norms reward familiar framing, and, most recently, LLMs, trained on the existing literature and inclined toward its most common phrasing, spread conventional wording at scale.\cite{filimonovic2026,sourati2026} On this view, LLMs are not the origin of the trend but are plausibly its latest mechanism, speeding up a homogenization already under way; separating these forces is a task for future work. 

Our primary measure captures semantic and linguistic similarity rather than conceptual novelty. Additionally, we conducted an analysis in terms of the main idea in a passage, providing evidence that the increase also extends to pairs expressing essentially the same scientific proposition. Nevertheless, this analysis is limited to ideas expressed at the passage level, rather than broader ideas such as the main contribution of an entire paper. Addressing that question would require different methods and is left for future work. Still, declining textual distinctiveness may make new contributions harder to identify. Across two independent databases, the scientific literature increasingly contained more of the same; broader and causal studies are needed to explain why.

These findings also suggest a practical application. If the scientific system can disadvantage novel work,\cite{wang2017} tools that make originality visible could act as a counterweight, helping authors, reviewers, editors and funders see how a contribution relates to what already exists, and giving distinctiveness a way to be recognized rather than overlooked. We would caution against turning any such measure into a target: a distinctiveness score optimized directly would reward unusual wording over genuine insight, the very confusion our findings warn against, and would be easy to game. 
Instead, such tools could highlight originality as one input among many, helping ensure that a growing scientific literature remains original rather than becoming more of the same.

\section*{Methods}

\subsection*{CORE corpus and paper selection}

Open access research papers were retrieved through CORE from May to July 2026.\cite{core} The exact search queries were \emph{astrophysics, bioinformatics, chemistry, climate change, machine learning, medicine, psychology and quantum computing}. The final eight queries were fixed before the analysis and chosen to span diverse scientific domains. Two additional queries, \emph{cybersecurity} and \emph{Alzheimer's disease}, were excluded because they did not yield sufficient eligible papers to meet the fixed target and source caps across all years. We used 2015--2025 to provide a long baseline and include the most recent available complete years.

 For each query and each target year from 2015 to 2025, the pipeline retained the first 300 eligible records in CORE result order as target papers. 
 CORE query results were ranked using Azure AI Search (personal communication, Catherine Kuliavets, Community and Relationship Manager at CORE). For text queries, Azure AI Search ranks results using the BM25 algorithm.\cite{azuresearch}
 A target published in year $y=2015,\ldots,2025$ was searched against source papers from years $y-1$ through $y-5$. We used five years as a fixed operational definition of recent literature.
  Source pools consisted of the first 600 eligible records in CORE result order from each source year, for a total of 3,000 papers per target year and query. The eight queries and all years reached that cap.
Hence, the analysis contained 3,300 target papers per query and 26,400 overall. The queries overlapped, yielding 26,059 unique CORE paper IDs; 334 records appeared in more than one query and were analyzed separately within each query.

\paragraph{Paper Filtration.} CORE records were processed using a cleaning pipeline. Filters based on section headings and text patterns removed detected reference sections, affiliations, repository text, and licensing and funding boilerplate. Records were eligible if they had a publication year, title, English full text, and at least 1,000 \emph{tokens} after cleaning, where a token was defined as a contiguous sequence of letters or digits after converting letters to lowercase. Administrative and teaching materials, other non-papers records, and records with extraction failures were excluded when detected. Each cleaned paper was divided into overlapping 160 token chunks, with consecutive chunks beginning 140 tokens apart and overlapping by 20 tokens, to avoid missing detection near the chunk boundaries. Only chunks lying entirely between 15\% and 78\% of
the cleaned document's token positions were eligible,
for both source and target papers. Chunks in which more than 25\% of tokens consisted entirely of digits were excluded, and the first 80 qualifying chunks in document order were retained per paper.

\subsection*{Semantic similarity detection}

Text chunks were embedded into numerical vectors with the 768-dimensional Sentence Transformers \texttt{all-mpnet-base-v2} model.\cite{sbert,mpnet} For embedding vectors $u$ and $v$, cosine similarity was defined as
$
\operatorname{cos}(u,v)=\frac{u^\top v}{\lVert u\rVert_2\lVert v\rVert_2}.
$
Because all embeddings were normalized to unit Euclidean norm, this was equal to their dot product $u^\top v$. 
For each target chunk, we retained for the analysis the 100 most similar source chunks among those with cosine similarity at least 0.82.

\paragraph{Definition of a match.}{\em Terms} were distinct tokens of at least five characters after excluding predefined stopwords, generic scientific vocabulary, query keywords, tokens made only of numbers, and tokens with several digits shorter than eight characters. A pair consisting of a target chunk and a source chunk was considered a {\em match} only if it met all of the following conservative criteria: cosine similarity at least 0.84; at least six shared terms; term overlap at least 0.10; no exact shared sequence of five consecutive tokens; no shared author; the papers were not identified as duplicates or versions; and both chunks passed the predefined quality filters.
In more detail,
if $S_t$ and $S_s$ denote the non-empty sets of terms in the target and source chunks $t,s$, respectively, {\em term overlap} was defined as
$
\frac{|S_t\cap S_s|}{\min(|S_t|,|S_s|)}.
$ The cosine threshold of 0.84 was chosen to prioritize precision, and robustness was assessed across thresholds from 0.82 to 0.90. Two synthetic examples illustrating matches are shown in the Extended Data~\ref{app:synthetic-matches}.

\paragraph{Definition of coverage.} For target paper $i$, let $T_i$ be the number of token occurrences in the entire cleaned paper, including unscreened portions, and number them from $1$ to $T_i$ in the order in which they appear. Each match involves a target chunk containing consecutive tokens $a,\ldots,b-1$, possibly with repetitions, represented by the set $\{a,\ldots,b-1\}$. Let $\mathcal I_i$ be the collection of such sets corresponding to matched target chunks in paper $i$. {\em Semantic similarity coverage} is the fraction of token occurrences contained in at least one matched chunk:
$
r_i=\frac{\left|\bigcup_{I\in\mathcal I_i} I\right|}{T_i}.
$
Thus, token occurrences contained in overlapping matched chunks were counted only once.
For each query and year, our primary measure was the mean of $r_i$ across the 300 target papers. We also examined the share of papers with at least one match, but this binary measure is somewhat more sensitive; therefore, mean $r_i$ was considered the more robust measure. Summaries gave each query equal weight. 

\subsection*{Controls and calibration}

\paragraph{synthetic null experiment.} We used the following controlled checks of the detector by constructing a benchmark containing 2,000 positive pairs created by rearranging word blocks within passages and 4,000 negative control pairs, 700 pairs from the same paper, 1,300 same-topic pairs and 2,000 unrelated pairs (Extended Data Fig.~\ref{ed:calibration}a,b). We evaluated the matching rule at the same threshold used in the main analysis.

Two additional diagnostics focused on a single query, \emph{bioinformatics} described next. Both semantic screens required cosine similarity $\geq 0.82$, at least 3 shared terms, term overlap $\geq 0.04$, and Jaccard similarity of 5-token and 8-token sequences was at most 0.08 and 0.03, respectively.

\paragraph{Semantic vs. lexical similarity.}
We compared semantic similarity against direct wording overlap (lexical overlap) to test whether the trend reflected copied language rather than broader semantic similarity (Extended Data Fig.~\ref{ed:lexical_semantic}a). Direct wording overlap was measured using a separate lexical matching procedure with the same target years, target cap and source pool size as the semantic analysis. The procedure first identified exact shared sequences of eight consecutive tokens and used them as starting points for detecting longer matching passages. Candidate passages required at least three such shared sequences. Candidate passages separated by at most 40 tokens were merged, and retained passages were required to span between 80 and 1,500 tokens. Retained passages were also required to have local token overlap of at least 0.25. 
The resulting matched passages were summarized as target token coverage using the same denominator as the semantic similarity measure.

\paragraph{Topic null.} We conducted a topic null experiment in which 300 \emph{bioinformatics} target papers from each year in 2015--2025 were compared with 3,000 \emph{climate change} source papers, comprising 600 papers from each of the preceding five years (Extended Data Fig.~\ref{ed:lexical_semantic}b). The same broader semantic matching procedure described above was used.

\subsection*{Citation linkage in CORE}

The primary screen yielded 10,287 matched passage pairs observations,
corresponding to 6,758 distinct source and target paper pairs within
queries (6,749 distinct paper pairs across queries).
We queried the databases OpenAlex, Semantic Scholar, Crossref, OpenCitations and Europe PMC for evidence that the target paper cited the source paper.\cite{openalex,semanticscholar,crossref,europepmc,opencitation} Separately, we searched the CORE full text of each target paper for the complete source title after normalizing case, punctuation and whitespace to reduce the false-negative rate.  
A pair was classified as \emph{citation linked} if either the bibliographic services or the CORE title search returned positive evidence. It was classified as \emph{no detected citation linkage} if neither method returned positive evidence and at least one method completed with a negative result. It was classified as \emph{unresolved} if neither positive nor conclusive negative evidence was available. No detected linkage was not interpreted as proof that a citation was absent. 

\paragraph{Human verification audit.} We assessed the classification using a stratified sample of two pairs from every combination of eight queries and eleven target years, yielding 176 pairs. Citation evidence for these pairs was manually verified and compared against the automated results. Because the number of matched pairs differed across groups of query and year in the full dataset, the audit estimate was weighted by the number of pairs in each group.

\subsection*{Idea level analysis}

We performed a post hoc analysis to test whether the increase in semantic similarity was also reflected at the level of scientific ideas. The analysis was designed conservatively, prioritizing precision over recall in identifying passage pairs that expressed essentially the same scientific proposition.
For this analysis, we used at most the first 800 characters of each passage.
We first screened the matched passage pairs for substantive scientific content, to retain only matches containing substantive scientific content, as some non-scientific content could pass our automated filtering. Claude Opus 5 independently classified each unique passage as substantive or non-substantive. Non-substantive material included references, author and affiliation lists, metadata, etc. A matched pair was retained for the idea level analysis only when both passages were classified as substantive. This excluded 344 of the 10,287 detected matches, leaving 9,943 pairs classified as substantive.

We then reduced each passage to its main idea. Namely, each unique passage was independently summarized by Claude Sonnet 5 into a short statement of its central scientific idea, targeting at most 15 words with a strict limit of 20. Each passage was summarized without showing the model its paired passage. For pairs in which both passages were substantive, Claude Opus 5 then received only the two resulting summaries and classified their relationship as \emph{same}, \emph{similar}, \emph{different}, or \emph{uncertain}. The classifier was blinded to the original passage text and to any other metadata. We defined \emph{same} conservatively as near equivalence of the central scientific proposition of the passage. \emph{Similar} required a specific substantive scientific proposition that was central to both summaries but did not reach near equivalence; sharing only the same subject was insufficient to be called \emph{similar}. 

For the analysis reported in the main text, we focused on pairs classified as \emph{same}. Of the 9,943 substantive matched pairs, 614 (6.2\%) met this strict criterion. Temporal comparisons used the annual numbers of these pairs and the same period definitions as in the main analysis.

\subsection*{Statistical analysis}

As a common robustness period contrast, we averaged the 48 means of field--year pairs from 2015--2020 (14,400 records) and 40 field--year means from 2021--2025 (12,000 records), giving every field and year equal weight. The main analysis used the annual trajectory, and we additionally compared 2025 with the 2015--2020 mean as an endpoint effect size.
Confidence intervals were obtained by resampling with replacement the 300 target papers within each pair of field and year. As a sensitivity check, papers were divided into ten groups by their order in the CORE results and resampled within these groups, preserving the acquisition order distribution. We also repeated the 2015--2020 versus 2021--2025 comparison after omitting each query and each year in turn.

\paragraph{Permutation testing.}
We conducted an exact one-sided permutation test of the hypothesis that: annual similarity coverage was initially stable and then entered a sustained increasing trajectory. For each candidate onset year from 2018 to 2023, we compared the annual equal-field mean coverage with a trajectory that was constant before the candidate onset and increased linearly thereafter. The test statistic was the largest Pearson correlation obtained across the candidate onset years. The observed trajectory was best aligned with an onset in 2021. The observed maximum Pearson correlation was $0.9913$.

Under the null hypothesis: the chronological ordering of the 11 annual observations is unrelated to similarity coverage. The one-sided alternative was that the observations are unusually consistent with an initially stable level followed by a sustained increase. We exhaustively evaluated all $11!=39{,}916{,}800$ permutations of the year labels, keeping the eight field means for each year together. For each permutation, we again selected the best-fitting onset year from 2018 to 2023 and recomputed the test statistic. Of all permutations, 720 produced a statistic at least as large as that observed; these were exactly the $6!$ permutations differing only in the ordering of the 2015--2020 years. The resulting exact one-sided $P$ value was $720/11!=1.80\times10^{-5}$.

\subsection*{Europe PMC replication}

Europe PMC provides an independent full-text database for scientific papers.\cite{europepmc} We used three queries: \emph{machine learning, climate change}, and \emph{psychology}. A fourth query, \emph{quantum computing}, was excluded because the source pool did not reach 3,000 papers in earlier years.
The same setting as in CORE was used: for each query and target year from 2015 to 2025, we retained 300 eligible target papers. Each target year used a balanced pool of 600 eligible source papers from each of the preceding five years, giving 3,000 source papers per query and target year.

Full texts were obtained in Journal Article Tag Suite (JATS) XML format and processed using the same cleaning, chunking and similarity pipeline as in the CORE analysis. All preprocessing, candidate retrieval, matching thresholds and filtering criteria were unchanged from the CORE analysis. Citation linkage was assessed separately using the reference information available through Europe PMC, and pairs whose citation status could not be resolved were classified as unresolved.
Annual summaries and confidence intervals were computed as in the CORE analysis, with equal weighting across queries.
Overlap between the CORE and Europe PMC samples for the three shared
queries was small. Of 9,900 target records in each database, 162
(1.64\%; 162 distinct papers) appeared in both. Among source records,
464 were shared (1.66\% of CORE sources and 1.72\% of Europe PMC
sources; 464 distinct papers).

\subsection*{Robustness}

\paragraph{Composition robustness.} We tested whether changes in target paper composition could explain the trend using several analyses. To account for differences in retrieval position in the database and paper length, we divided papers in each query into three retrieval order groups and five paper length groups, creating 15 combinations. We then adjusted every year to have the same proportions across these combinations as the full 2015--2025 sample for that query. 
We also restricted the analysis to DOI prefixes (the first part of a DOI, often indicating the publisher) represented by at least 10 papers in 2015--2020 and 5 papers in 2021--2025, and present in at least four baseline years and two later years. We additionally adjusted the two periods to have the same distribution of these persistent DOI prefixes within each query. For the above robustness checks, see Extended data Fig.~\ref{ed:composition_controls}.

\paragraph{Topic composition and publication lag.}
To assess changes in topic composition, each paper from 2021--2025 was compared within its query with the five papers from 2015--2020 having the most similar titles. Similarity was measured using term frequency--inverse document frequency (TF--IDF) based on single words and pairs of consecutive words, with English stop words removed; matching was performed with replacement.
To test whether differences in the ages of the source papers could explain the trend, we repeated the analysis separately at each {\em publication lag} (the difference between the target and source publication years), retaining only matches to source papers published exactly one, two, three, four or five years before each target paper (Fig.~\ref{fig:main}d). We also repeated the analysis across cosine similarity thresholds from 0.82 to 0.90 and separately across chunk sizes 80, 180, 320, and 480 (Extended Data Fig.~\ref{ed:chunk_robustness}d).

\paragraph{Other Robustness checks.}
To assess whether overlap between the target and source samples affected the result, we additionally restricted matches to source papers that were never analyzed as targets within the same query.
To rule out that the trend was caused only by a small number of repeatedly reused sources, we repeated the analysis after excluding the top 1\%, 5\% and 10\% of source papers ranked within each query by the number of distinct matched target papers. We separately excluded source passages matched to at least 2, 3, or 5 distinct target papers within the same query.
To test whether the trend depended on particular subsets of source papers or passages, we assigned them separately to reproducible groups using SHA-256 hashes of identifiers depending on the query and the paper ID, and repeated the analysis while omitting each group in turn.
This analysis used 40 groups, with sensitivity analyses using 20 and 80 groups. To control for changes in the number of passages,
we estimated expected coverage under uniform random
subsampling to fixed source and target passage counts,
using the recorded matches without rerunning the search.

\paragraph*{arXiv analysis}
We conducted a complementary analysis using arXiv papers from a narrow computer science domain. Eligible papers were assigned to all three of \emph{cs.CL}, \emph{cs.LG}, and \emph{cs.AI}, without a restriction on the primary category. For each target year from 2015 to 2025, up to 300 eligible target papers were analyzed, using the same matching pipeline as in the CORE analysis. Each target sample was compared with all available eligible source papers from the preceding five years; unlike the CORE analysis, the number of source papers was therefore allowed to grow with the available literature.

\paragraph{Lexical filters.}
To determine whether standardized methods language accounted for the trend, we repeated the analysis after excluding matched target chunks classified as technical methods using automated section labels based on headings, position, and lexical cues.
As an additional sensitivity analysis, we repeated the \emph{machine learning} analysis on 250 target papers per year and up to 1,500 source papers per target year, with a generic ML content filter enabled and disabled while holding the paper selection and all other settings fixed. Unlike the main analysis, this sensitivity analysis allowed shared exact five-grams and nonzero author-overlap scores. This filter was kept for the 8 main runs and an additional sensitivity run on \emph{bioinformatics} confirmed that this filter has a negligible effect on non machine learning queries.

\paragraph{Crowding of the embedding space.}
We tested whether the trend could reflect increasing crowding of the embedding space. To this end, we performed an additional diagnostic in bioinformatics. For each year from 2015 to 2025, we sampled 4,000 target passage embedding vectors without replacement, repeated this independently five times, and excluded comparisons between passages from the same paper. We measured the fraction of inter-paper passage pairs with cosine similarity at least 0.84, the same threshold used in the main screen. Annual values were averaged across the five repeats. See Extended data Fig.~\ref{ed:crowding}.

\subsection*{Limitations}

The CORE study and Europe PMC replication use fixed literal queries and ranked retrieval rather than random uniform samples of all scientific papers. Database coverage, rankings and the availability of full text may also change over time.\cite{visser2021} Our robustness analyses test sensitivity within the retrieved samples, but they cannot account for relevant papers absent from the retrieved results or make the samples representative of science as a whole. Composition controls are limited to observed metadata. 

The detector cannot identify similarity to unavailable, unsampled, older or strongly transformed sources. The 80-chunk cap limits coverage of long papers, and candidate retrieval retained only the 100 most similar source chunks for each target chunk. Performance on the constructed benchmark may not fully represent naturally occurring scientific text. Citation resolution is incomplete, so no detected citation linkage does not fully guarantee that a citation was absent.
The arXiv analysis differed from the CORE analysis 
as arXiv papers were not selected through ranked database query results. 
It should not be interpreted as a replication of the CORE or Europe PMC experiments.
Our primary measure captures textual similarity rather than conceptual novelty; the idea-level analysis is limited to detected passage matches and does not assess the novelty of the main idea of a paper. 
Finally, the observed trajectory does not identify a unique cause and cannot establish that LLM use directly caused the increase.

\subsection*{Declarations}

\paragraph{AI Usage.}
ChatGPT was used to assist with code development and drafting the manuscript. Claude API was used for the idea level experiments. 
AI did not determine the study endpoint, acquire the original CORE
corpus or determine primary corpus eligibility. LLM classifications
were used only in the passage idea analysis. The authors verified the analyses and are responsible for all claims and conclusions.

\paragraph{Funding.} IDA is supported by grant NSF DMS-2031883 and Vannevar Bush Faculty Fellowship ONR-N00014-20-1-2826 (PI Mossel). EM is partially supported by NSF DMS-2031883, Vannevar Bush Faculty Fellowship ONR-N00014-20-1-2826, MURI N000142412742, and a Simons Investigator Award.

\appendix

\newpage
\section{Extended Data}

\begin{figure}[H]
\centering

\vspace{0.5em}

\noindent
\begin{minipage}{0.035\linewidth}
    \centering
    \rotatebox{90}{\scriptsize Target text matched to prior sources (\%)}
\end{minipage}%
\begin{minipage}{0.955\linewidth}
    \centering

    \begin{minipage}[t]{0.24\linewidth}
        \raggedright\textbf{a}\par
        \vspace{0.05em}
        \centering
        \includegraphics[width=\linewidth]{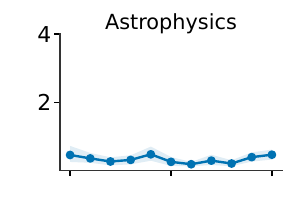}
    \end{minipage}\hfill
    \begin{minipage}[t]{0.24\linewidth}
        \raggedright\textbf{b}\par
        \vspace{0.05em}
        \centering
        \includegraphics[width=\linewidth]{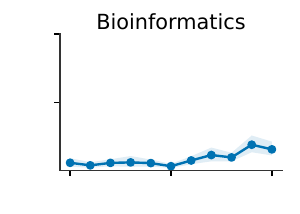}
    \end{minipage}\hfill
    \begin{minipage}[t]{0.24\linewidth}
        \raggedright\textbf{c}\par
        \vspace{0.05em}
        \centering
        \includegraphics[width=\linewidth]{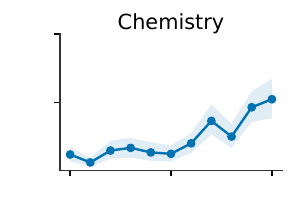}
    \end{minipage}\hfill
    \begin{minipage}[t]{0.24\linewidth}
        \raggedright\textbf{d}\par
        \vspace{0.05em}
        \centering
        \includegraphics[width=\linewidth]{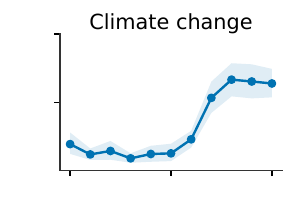}
    \end{minipage}

    \vspace{0.15em}

    \begin{minipage}[t]{0.24\linewidth}
        \raggedright\textbf{e}\par
        \vspace{0.05em}
        \centering
        \includegraphics[width=\linewidth]{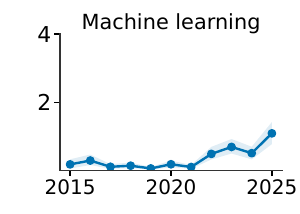}
    \end{minipage}\hfill
    \begin{minipage}[t]{0.24\linewidth}
        \raggedright\textbf{f}\par
        \vspace{0.05em}
        \centering
        \includegraphics[width=\linewidth]{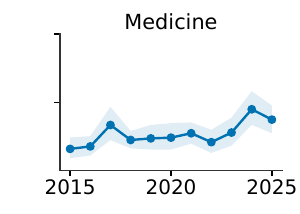}
    \end{minipage}\hfill
    \begin{minipage}[t]{0.24\linewidth}
        \raggedright\textbf{g}\par
        \vspace{0.05em}
        \centering
        \includegraphics[width=\linewidth]{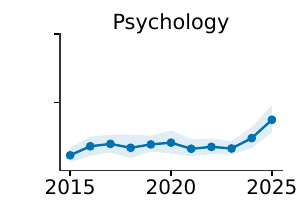}
    \end{minipage}\hfill
    \begin{minipage}[t]{0.24\linewidth}
        \raggedright\textbf{h}\par
        \vspace{0.05em}
        \centering
        \includegraphics[width=\linewidth]{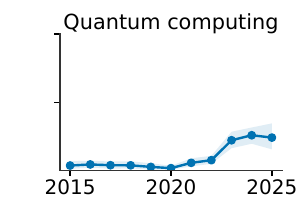}
    \end{minipage}

    \vspace{0.15em}

    {\scriptsize Target year}
\end{minipage}

\vspace{0.3em}

\caption{\textbf{Semantic similarity coverage across scientific fields.}
\textbf{a--h}, Annual semantic similarity coverage trends for the eight individual queries.
Lines show the mean percentage of target text matched to prior sources, and shaded bands indicate 95\% bootstrap confidence intervals over target papers (1,000 bootstrap resamples).
Each field contributes equally to the aggregate in Fig~\ref{fig:main}a.}

\label{fig:main_trends}
\end{figure}

\begin{figure}[H]
\centering

\noindent
\begin{minipage}{0.035\linewidth}
    \centering
    \rotatebox{90}{\scriptsize Share of papers with $\geq 1$ matched passage (\%)}
\end{minipage}%
\begin{minipage}{0.955\linewidth}
    \centering

    \begin{minipage}[t]{0.24\linewidth}
        \raggedright\textbf{a}\par
        \vspace{0.05em}
        \centering
        \includegraphics[width=\linewidth]{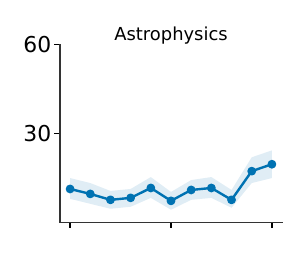}
    \end{minipage}\hfill
    \begin{minipage}[t]{0.24\linewidth}
        \raggedright\textbf{b}\par
        \vspace{0.05em}
        \centering
        \includegraphics[width=\linewidth]{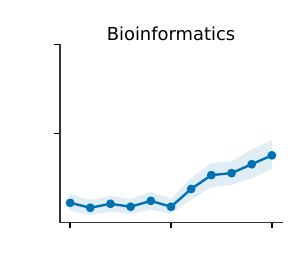}
    \end{minipage}\hfill
    \begin{minipage}[t]{0.24\linewidth}
        \raggedright\textbf{c}\par
        \vspace{0.05em}
        \centering
        \includegraphics[width=\linewidth]{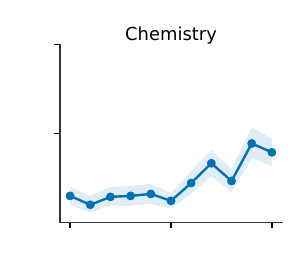}
    \end{minipage}\hfill
    \begin{minipage}[t]{0.24\linewidth}
        \raggedright\textbf{d}\par
        \vspace{0.05em}
        \centering
        \includegraphics[width=\linewidth]{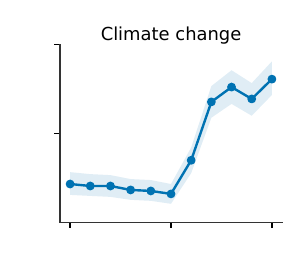}
    \end{minipage}

    \vspace{-0.25em}

    \begin{minipage}[t]{0.24\linewidth}
        \raggedright\textbf{e}\par
        \vspace{0.05em}
        \centering
        \includegraphics[width=\linewidth]{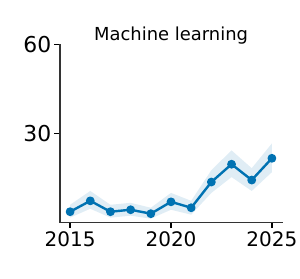}
    \end{minipage}\hfill
    \begin{minipage}[t]{0.24\linewidth}
        \raggedright\textbf{f}\par
        \vspace{0.05em}
        \centering
        \includegraphics[width=\linewidth]{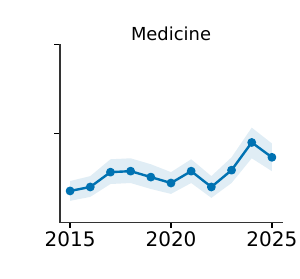}
    \end{minipage}\hfill
    \begin{minipage}[t]{0.24\linewidth}
        \raggedright\textbf{g}\par
        \vspace{0.05em}
        \centering
        \includegraphics[width=\linewidth]{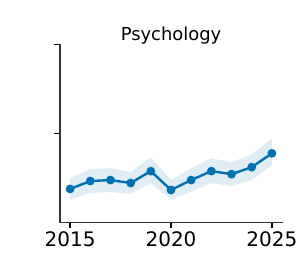}
    \end{minipage}\hfill
    \begin{minipage}[t]{0.24\linewidth}
        \raggedright\textbf{h}\par
        \vspace{0.05em}
        \centering
        \includegraphics[width=\linewidth]{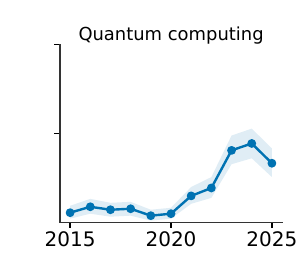}
    \end{minipage}

    \vspace{-0.15em}

    {\scriptsize Target year}
\end{minipage}

\vspace{-0.3em}

\caption{\textbf{Share of papers with at least one match across scientific fields.}
\textbf{a--h}, Annual share of papers containing at least one matched passage. 
Shaded bands indicate 95\% bootstrap confidence intervals over target papers (1,000 bootstrap resamples). Each field contributes equally to the aggregate in Fig~\ref{fig:main}b.}

\label{fig:field_prevalence}
\end{figure}

\begin{figure*}[t]
    \centering
    \begin{minipage}[t]{0.48\textwidth}
        \textbf{a}\par\vspace{2pt}
        \centering
        \includegraphics[
            width=\linewidth,
            height=0.21\textheight,
            keepaspectratio
        ]{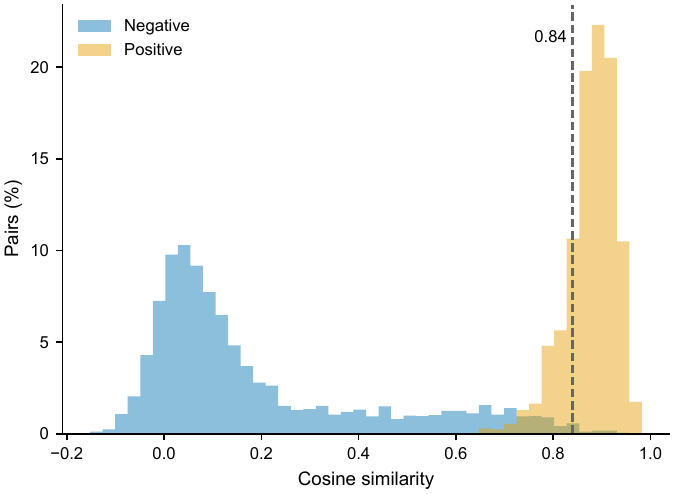}
    \end{minipage}
    \hfill
    \begin{minipage}[t]{0.48\textwidth}
        \textbf{b}\par\vspace{2pt}
        \centering
        \includegraphics[
            width=\linewidth,
            height=0.21\textheight,
            keepaspectratio
        ]{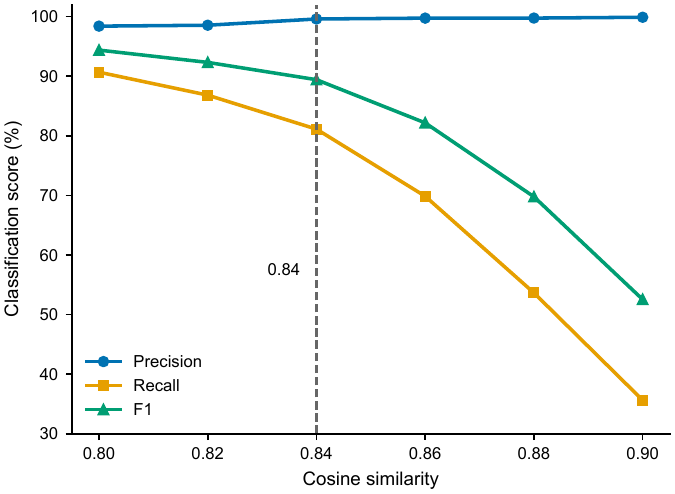}
    \end{minipage}

    \caption{\textbf{Validation and calibration of the semantic similarity detector.}
    \textbf{a}, Distribution of cosine similarity for 2,000 constructed positive pairs and 4,000 negative pairs, normalized separately within each class. The dashed line marks the frozen cosine threshold of 0.84.
    \textbf{b}, Precision, recall and F1 of the matching rule across cosine thresholds. At the frozen threshold of 0.84, precision was 99.6\%, recall was 81.1\% and F1 was 89.4\%.}
    \label{ed:calibration}
\end{figure*}

\begin{figure*}[t]
    \centering

    \vspace{-0.5em}
    \includegraphics[width=0.58\textwidth]{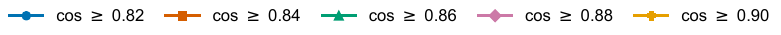}
    \par\vspace{-1.0em}

    \begin{minipage}[t]{0.24\textwidth}
        \vspace{0pt}
        \textbf{a}\par\vspace{0.25em}
        \includegraphics[width=\linewidth]{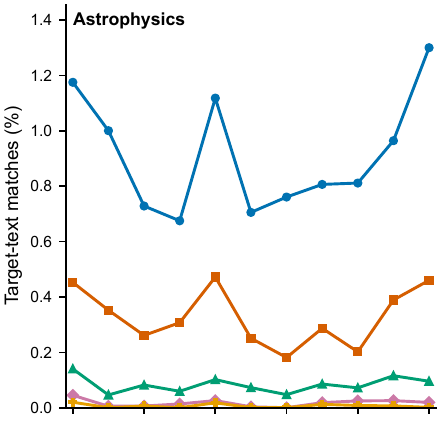}
    \end{minipage}\hfill
    \begin{minipage}[t]{0.24\textwidth}
        \vspace{0pt}
        \textbf{b}\par\vspace{0.25em}
        \includegraphics[width=\linewidth]{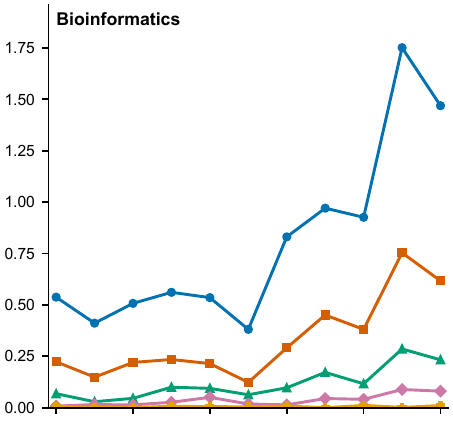}
    \end{minipage}\hfill
    \begin{minipage}[t]{0.24\textwidth}
        \vspace{0pt}
        \textbf{c}\par\vspace{0.25em}
        \includegraphics[width=\linewidth]{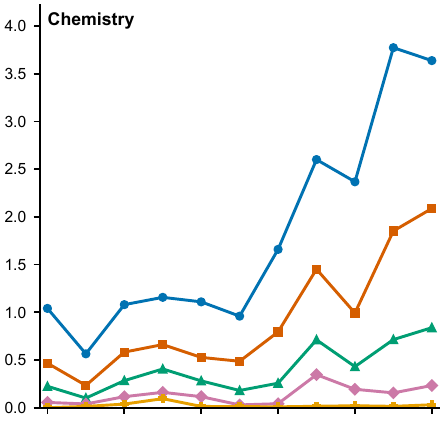}
    \end{minipage}\hfill
    \begin{minipage}[t]{0.24\textwidth}
        \vspace{0pt}
        \textbf{d}\par\vspace{0.25em}
        \includegraphics[width=\linewidth]{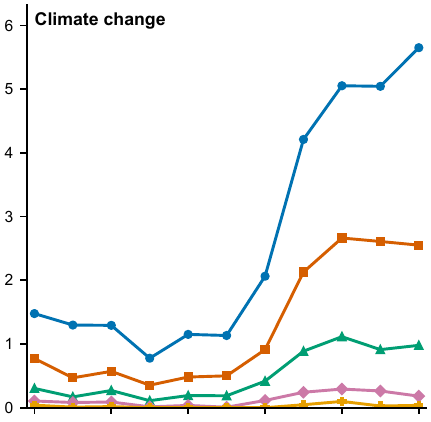}
    \end{minipage}

    \par\vspace{0.9em}

    \begin{minipage}[t]{0.24\textwidth}
        \vspace{0pt}
        \textbf{e}\par\vspace{0.25em}
        \includegraphics[width=\linewidth]{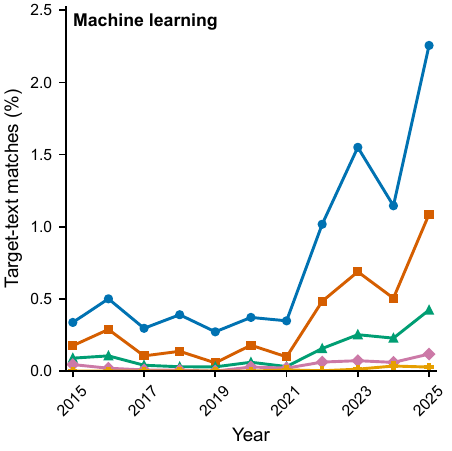}
    \end{minipage}\hfill
    \begin{minipage}[t]{0.24\textwidth}
        \vspace{0pt}
        \textbf{f}\par\vspace{0.25em}
        \includegraphics[width=\linewidth]{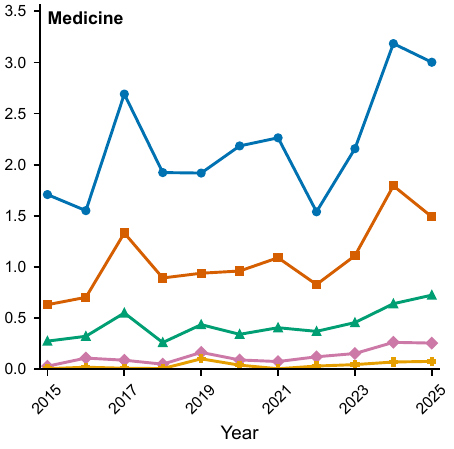}
    \end{minipage}\hfill
    \begin{minipage}[t]{0.24\textwidth}
        \vspace{0pt}
        \textbf{g}\par\vspace{0.25em}
        \includegraphics[width=\linewidth]{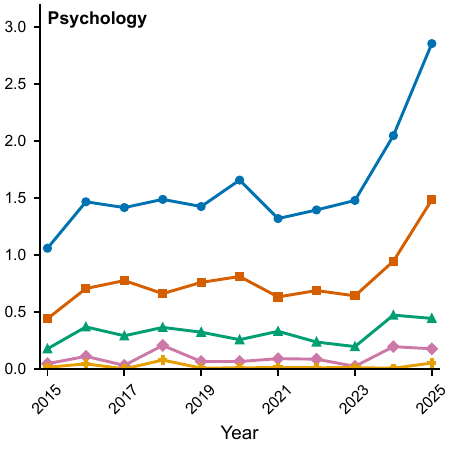}
    \end{minipage}\hfill
    \begin{minipage}[t]{0.24\textwidth}
        \vspace{0pt}
        \textbf{h}\par\vspace{0.25em}
        \includegraphics[width=\linewidth]{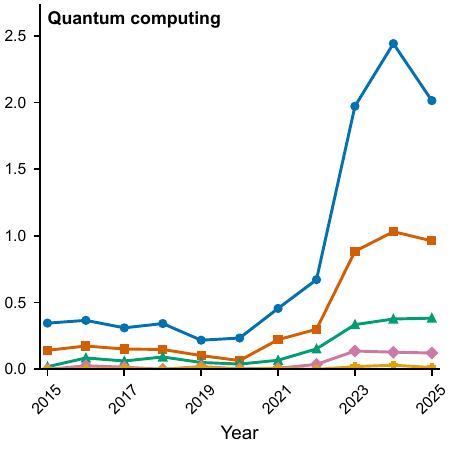}
    \end{minipage}

    \caption{\textbf{Robustness of the temporal trend to the cosine similarity threshold across fields.}
    Annual target text match coverage is shown for each of the eight fields while varying the cosine similarity threshold from 0.82 to 0.90. The temporal increase persists across most fields and threshold choices. Y-axis ranges vary across fields to make threshold sensitivity visible.}

    \label{ed:threshold_sensitivity}
\end{figure*}

\begin{figure*}[t]
    \centering

    \vspace{-1.5em}
    \includegraphics[width=0.43\textwidth]{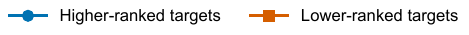}
    \par\vspace{-1.0em}

    \begin{minipage}[t]{0.24\textwidth}
        \vspace{0pt}
        \textbf{a}\par\vspace{0.25em}
        \includegraphics[width=\linewidth]{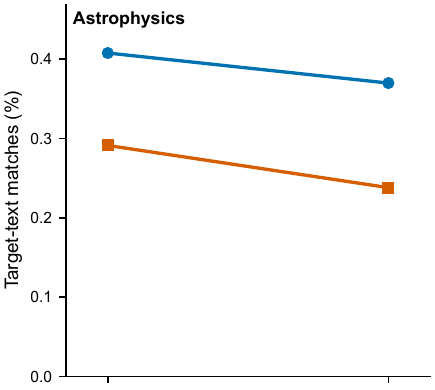}
    \end{minipage}\hfill
    \begin{minipage}[t]{0.24\textwidth}
        \vspace{0pt}
        \textbf{b}\par\vspace{0.25em}
        \includegraphics[width=\linewidth]{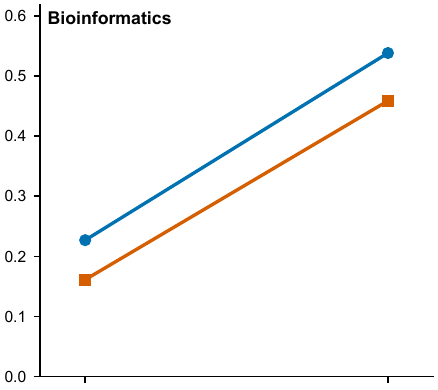}
    \end{minipage}\hfill
    \begin{minipage}[t]{0.24\textwidth}
        \vspace{0pt}
        \textbf{c}\par\vspace{0.25em}
        \includegraphics[width=\linewidth]{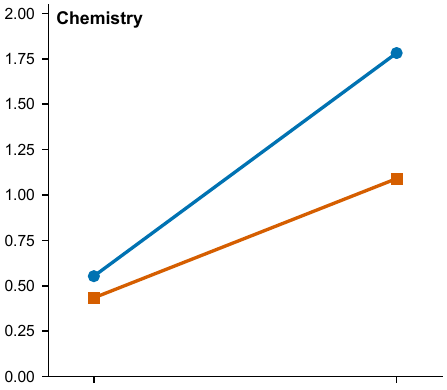}
    \end{minipage}\hfill
    \begin{minipage}[t]{0.24\textwidth}
        \vspace{0pt}
        \textbf{d}\par\vspace{0.25em}
        \includegraphics[width=\linewidth]{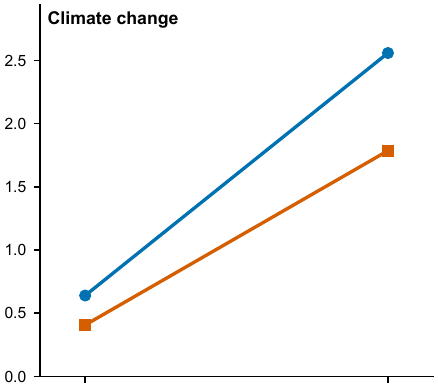}
    \end{minipage}

    \par\vspace{1.0em}

    \begin{minipage}[t]{0.24\textwidth}
        \vspace{0pt}
        \textbf{e}\par\vspace{0.25em}
        \includegraphics[width=\linewidth]{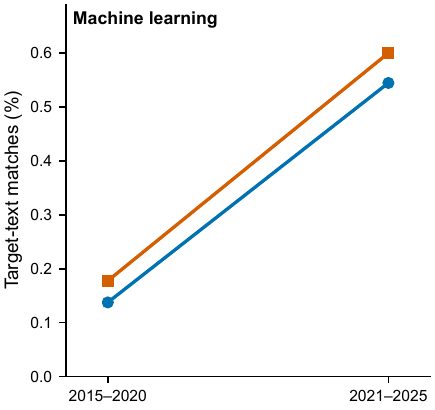}
    \end{minipage}\hfill
    \begin{minipage}[t]{0.24\textwidth}
        \vspace{0pt}
        \textbf{f}\par\vspace{0.25em}
        \includegraphics[width=\linewidth]{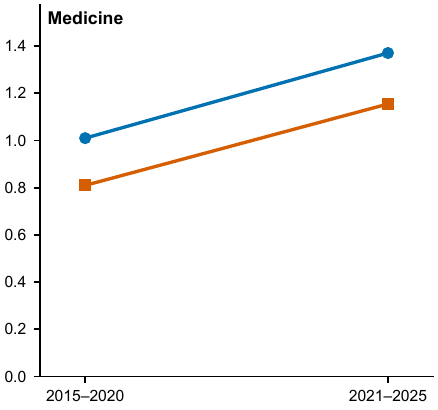}
    \end{minipage}\hfill
    \begin{minipage}[t]{0.24\textwidth}
        \vspace{0pt}
        \textbf{g}\par\vspace{0.25em}
        \includegraphics[width=\linewidth]{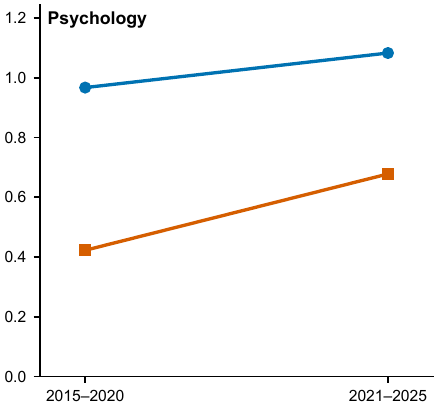}
    \end{minipage}\hfill
    \begin{minipage}[t]{0.24\textwidth}
        \vspace{0pt}
        \textbf{h}\par\vspace{0.25em}
        \includegraphics[width=\linewidth]{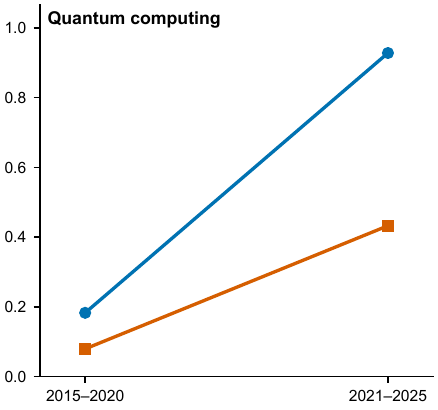}
    \end{minipage}

    \caption{\textbf{Robustness of the temporal increase to target rank within the sampled CORE results.}
    Within each field and year, the 300 sampled target papers were divided by CORE query rank into 150 higher-ranked and 150 lower-ranked targets. Mean target-text match coverage is shown for 2015--2020 and 2021--2025 for both groups across all eight fields. The increase is present in both higher and lower ranked targets except for \emph{astrophysics}, indicating that the temporal trend is not confined to the highest ranked sampled query results. Y-axis ranges vary across fields for visibility.}
    \label{ed:rank_period_robustness}
\end{figure*}

\begin{figure*}[t]

    \includegraphics[width=0.60\textwidth]{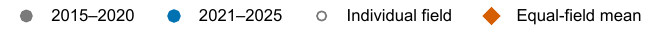}
    \par\vspace{-1.0em}

    \begin{minipage}[t]{0.53\textwidth}
        \vspace{0pt}
        \textbf{a}\par\vspace{0.25em}
        \includegraphics[width=\linewidth]{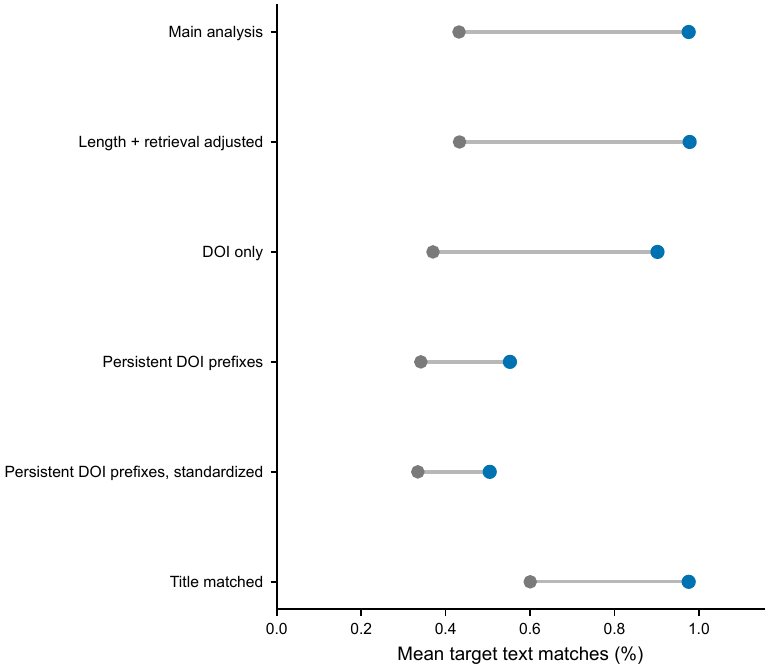}
    \end{minipage}\hfill
    \begin{minipage}[t]{0.45\textwidth}
        \vspace{0pt}
        \textbf{b}\par\vspace{0.25em}
        \includegraphics[width=\linewidth]{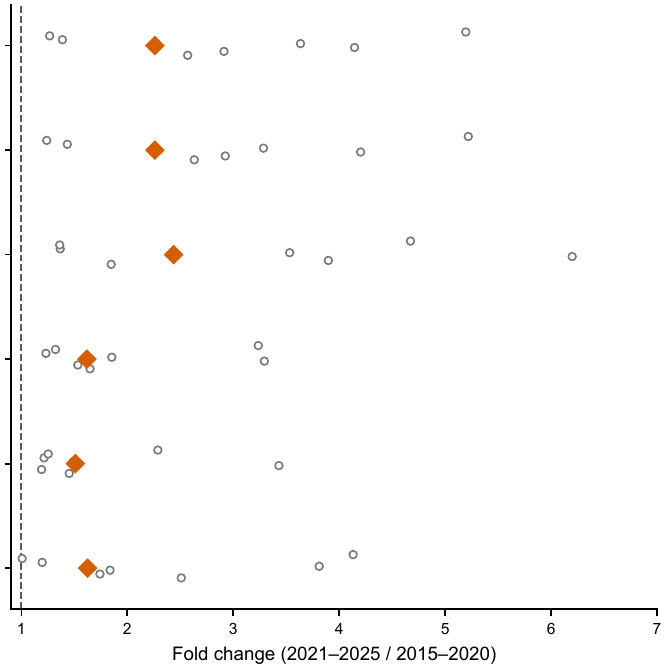}
    \end{minipage}

    \caption{\textbf{Robustness of the temporal increase to changes in the composition of the sampled literature.}
    \textbf{a}, Mean target text match coverage in 2015--2020 and 2021--2025 for the primary analysis and five composition controls: adjustment for paper length and retrieval order, restriction to papers with DOIs, restriction to persistent DOI prefixes, standardization across persistent DOI prefixes, and title matching.
    \textbf{b}, Corresponding fold changes for each individual field (open circles) and the equal-field mean (orange diamonds). The dashed vertical line indicates no change between periods.}
    \label{ed:composition_controls}
\end{figure*}

\begin{figure*}[t]
    \centering

    \begin{minipage}[t]{0.48\textwidth}
        \centering
        \includegraphics[width=\linewidth]{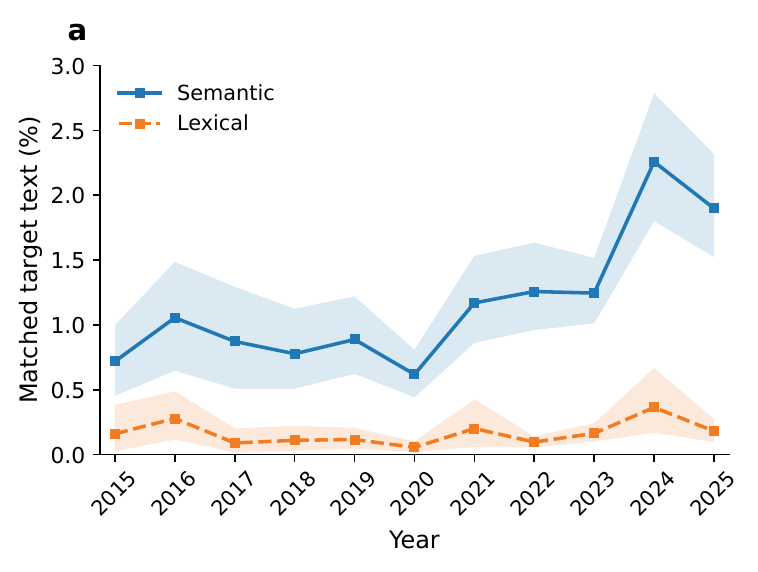}
    \end{minipage}
    \hfill
    \begin{minipage}[t]{0.48\textwidth}
        \centering
        \includegraphics[width=\linewidth]{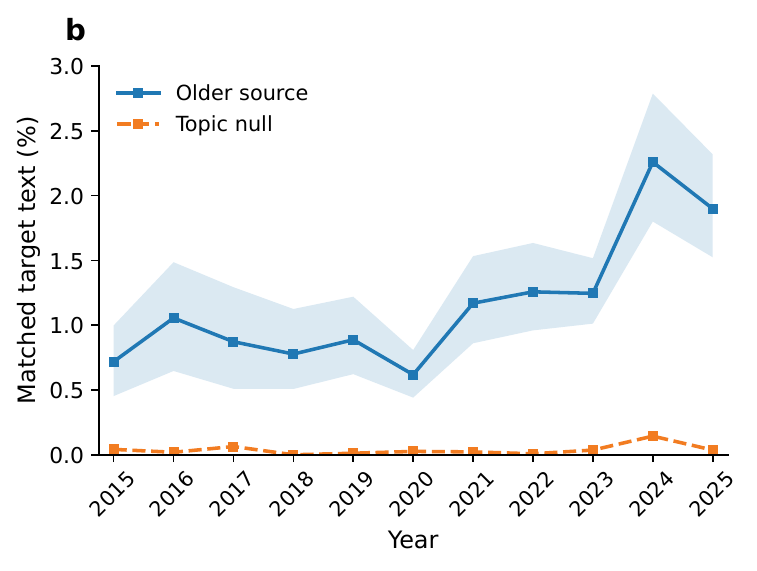}
    \end{minipage}

    \par\vspace{0.5em}

    \begin{minipage}[t]{0.48\textwidth}
        \centering
        \includegraphics[width=\linewidth]{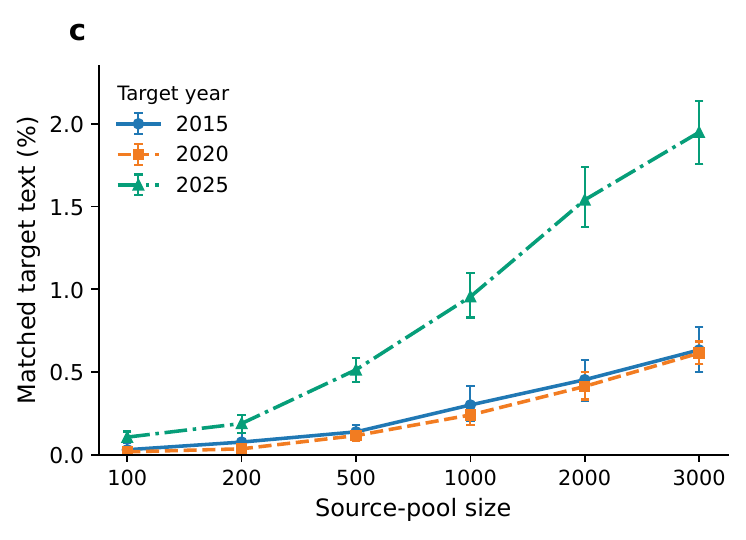}
    \end{minipage}
    \hfill
    \begin{minipage}[t]{0.48\textwidth}
        \centering
        \includegraphics[width=\linewidth]{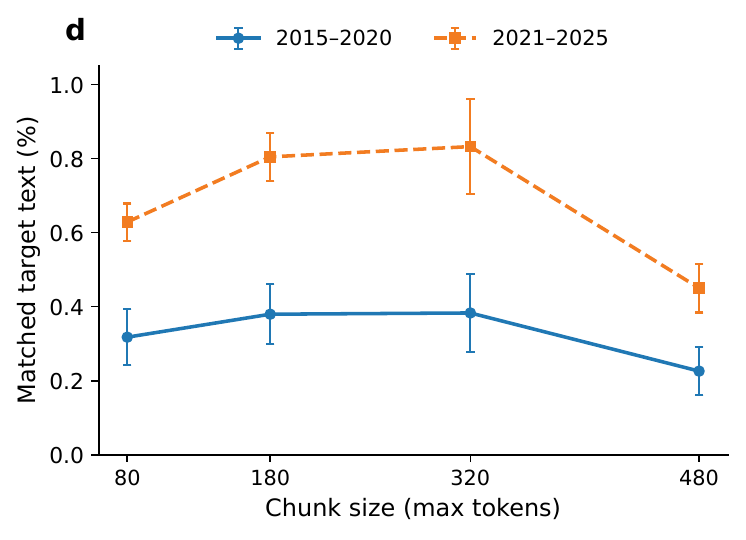}
    \end{minipage}

    \caption{\textbf{Validation and robustness of the temporal increase in semantic similarity.}
    \textbf{a}, Semantic similarity compared with direct lexical overlap in bioinformatics.
    \textbf{b}, Semantic similarity against prior same-topic papers compared with an unrelated-topic source pool. Shaded bands in a and b indicate 95\% bootstrap confidence intervals over target papers.
    \textbf{c}, Mean target text coverage in 2015, 2020 and 2025 across source-pool sizes. Each point averages 10 repeats using 100 target papers; error bars indicate 95\% bootstrap confidence intervals over repeats.
    \textbf{d}, Mean target text coverage across maximum passage chunk sizes for 2015--2020 and 2021--2025; error bars indicate $\pm1$ standard error across annual means (6 and 5 years, respectively).}

\label{ed:lexical_semantic}
\label{ed:topic_null}
\label{ed:pool_sweep}
\label{ed:chunk_robustness}
\end{figure*}

\begin{figure}[t]
    \centering

    \begin{minipage}[t]{0.48\linewidth}
        \textbf{a}\par\vspace{1mm}
        \centering
        \includegraphics[width=\linewidth]{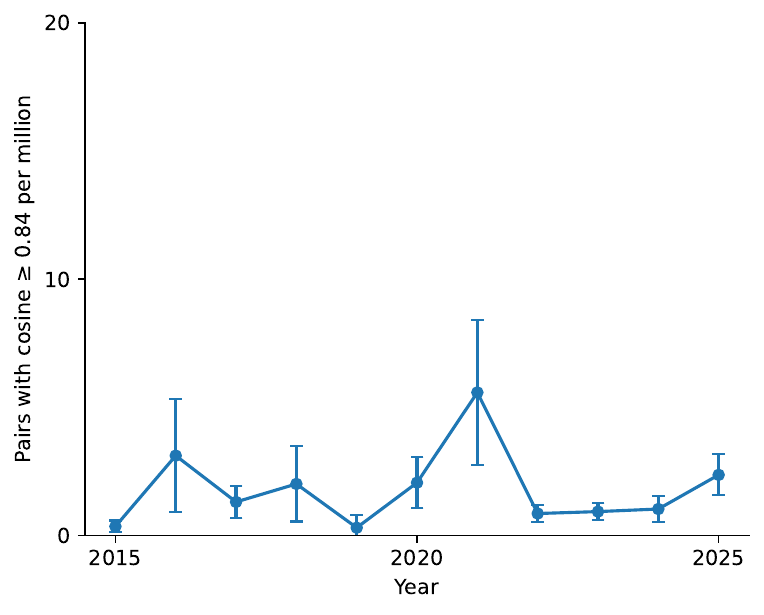}
    \end{minipage}
    \hfill
    \begin{minipage}[t]{0.48\linewidth}
        \textbf{b}\par\vspace{1mm}
        \centering
        \includegraphics[width=\linewidth]{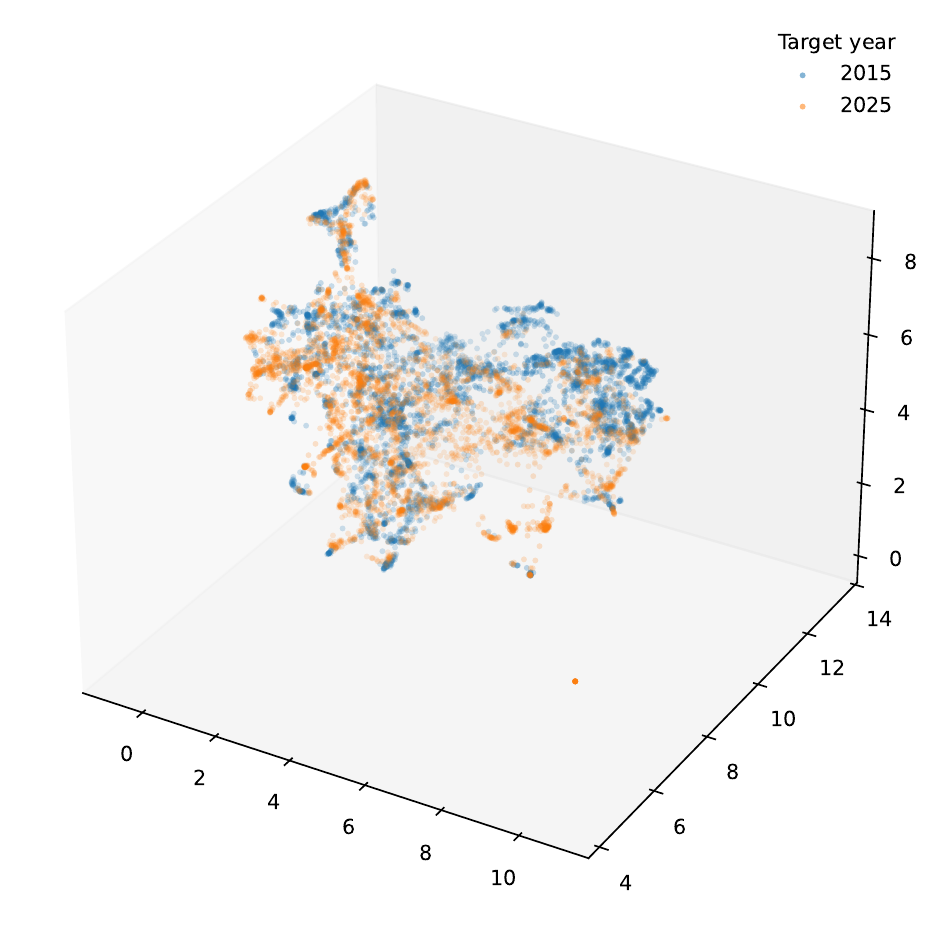}
    \end{minipage}

    \caption{\textbf{The increase in semantic similarity is not explained by increased crowding of the embedding space.}
    \textbf{a}, Frequency of inter-paper passage pairs exceeding the semantic similarity threshold across years.
    \textbf{b}, 
    a joint 3D projection of sampled passage embeddings from 2015 and 2025.}

    \label{ed:crowding}
\end{figure}

\clearpage
\nolinenumbers
\newcommand{\SameIdea}{\textcolor{teal!70!black}{\textbf{SAME}}}
\newcommand{\SimilarIdea}{\textcolor{orange!85!black}{\textbf{SIMILAR}}}
\newcommand{\DifferentIdea}{\textcolor{red!70!black}{\textbf{DIFFERENT}}}

\setlength{\columnsep}{1.2em}

\clearpage
\nolinenumbers

\setlength{\columnsep}{1.2em}

\clearpage
\nolinenumbers
\setlength{\columnsep}{1.2em}

\clearpage
\nolinenumbers
\setlength{\columnsep}{1.2em}

\clearpage
\nolinenumbers
\setlength{\columnsep}{1.2em}

\clearpage
\nolinenumbers
\setlength{\columnsep}{1.2em}

\section{Illustrative synthetic semantic matches}
\label{app:synthetic-matches}

{\small
We constructed two synthetic passage pairs to illustrate the type of
semantic match detected by our pipeline. The passages were not drawn from
the historical corpus and were not used for statistical inference. Each
passage is approximately 160 tokens after the same cleaning and tokenization
used in the main analysis. Both pairs satisfy the applicable frozen matching
rule: cosine similarity at least 0.84, at least six shared content-specific
terms, specific-term overlap at least 0.10, no shared exact five-token
sequence, and all passage-level quality filters. Paper-level author,
citation, duplicate and version checks are not applicable to these
standalone synthetic passages. Boldface marks selected shared
content-bearing terms.
}

\vspace{1.0em}

{\small
\textbf{Example 1:}
Cosine = \textbf{0.851681}, 
Specific-term overlap = \textbf{0.163043}, 
Shared terms = \textbf{15}, Idea: \textbf{SAME}
}

\vspace{0.25em}

\begin{paracol}{2}

{\footnotesize\textbf{SYNTHETIC SOURCE}}

\switchcolumn

{\footnotesize\textbf{SYNTHETIC TARGET}}

\end{paracol}

\vspace{0.1em}

\begin{paracol}{2}

{\scriptsize
``During a \textbf{transit}, \textbf{stellar} radiation crosses the thin
\textbf{atmosphere} surrounding an exoplanet before entering the telescope.
The apparent \textbf{radius} changes with \textbf{wavelength} because
different gases block different fractions of incoming flux. Features in the
resulting \textbf{spectrum} can therefore reveal \textbf{absorption} by
\textbf{molecular} species. Interpreting those signatures is difficult
because \textbf{opacity} also depends on pressure and thermal gradients,
while aerosols or \textbf{clouds} can suppress otherwise visible lines. A
\textbf{retrieval} procedure compares the measurement with many simulated
forward spectra and assigns support to alternative chemical mixtures and
vertical profiles. Repeated visits help separate \textbf{persistent}
\textbf{planetary} signatures from detector drift, \textbf{starspot}
\textbf{contamination}, and random noise. Even when individual abundances
remain uncertain, wide wavelength coverage can exclude some gaseous
scenarios and identify intervals where another observation would provide
the greatest information. Such evidence is indirect, but it permits
comparative studies of remote worlds that cannot yet be spatially resolved.
These estimates are useful for comparing chemistry across objects sampled
under otherwise similar geometric conditions.''
}

\switchcolumn

{\scriptsize
``Transmission spectroscopy uses a \textbf{transit} to examine material in
the \textbf{atmosphere} around a distant planet. As \textbf{stellar} photons
filter through its atmosphere, the effective \textbf{radius} varies with
\textbf{wavelength}, producing structure in the observed \textbf{spectrum}.
A localized \textbf{absorption} signal may be consistent with a
\textbf{molecular} absorber, although haze or \textbf{clouds} can create
degeneracies with gas abundance because both alter the effective
\textbf{opacity}. \textbf{Retrieval} analysis evaluates observed data
against modeled atmospheres spanning varied compositions, temperatures, and
height structures, then derives posterior ranges for several plausible
candidate physical states overall. Combining several observing epochs can
reduce sensitivity to \textbf{persistent} \textbf{planetary} signals being
confused with \textbf{starspot} \textbf{contamination} or instrumental
variation. Broad wavelength baselines are especially valuable because a
feature that is ambiguous in one channel may become distinguishable when
adjacent regions are included. The final result is usually a family of
plausible compositions rather than a unique description, and follow-up
measurements can target the portions carrying the most diagnostic leverage
for future observing campaigns.''
}

\end{paracol}

\begin{paracol}{2}

{\scriptsize
\textit{\textbf{Idea:}
Transit spectroscopy constrains exoplanet atmospheres by measuring
wavelength-dependent absorption in stellar light during transits.}
}

\switchcolumn

{\scriptsize
\textit{\textbf{Idea:}
Transit spectra constrain atmospheric composition by linking
wavelength-dependent radius changes to molecular absorption and opacity.}
}

\end{paracol}

\vspace{1.0em}

{\small
\textbf{Example 2:}
Cosine = \textbf{0.849445}, 
Specific-term overlap = \textbf{0.132530}, 
Shared terms = \textbf{11}, Idea: \textbf{SIMILAR}
}

\vspace{0.25em}

\begin{paracol}{2}

{\footnotesize\textbf{SYNTHETIC SOURCE}}

\switchcolumn

{\footnotesize\textbf{SYNTHETIC TARGET}}

\end{paracol}

\vspace{0.1em}

\begin{paracol}{2}

{\scriptsize
``A classifier can produce sharp \textbf{probabilities} even when its
reported \textbf{confidence} does not match empirical accuracy. Temperature
scaling is a post hoc \textbf{calibration} method that adjusts this mismatch
without retraining the \textbf{predictor}. After optimization is complete,
the \textbf{logits} are divided by a learned \textbf{temperature} before
normalization. The scalar is selected on a held-out validation partition
containing known labels, while all original network weights stay fixed.
Larger values soften the probabilities and can reduce systematic
\textbf{overconfidence} without altering the highest ranked output. This
adjustment is often effective when future cases resemble the
\textbf{distribution} used for validation. It is less reliable after major
covariate change, because one global adjustment cannot repair
input-dependent or label-specific errors. Richer calibrators can represent
more complex distortions, but they need additional fitting data and may
increase estimation \textbf{uncertainty}. For that reason, calibration
quality should be checked separately from discrimination whenever predictive
scores are used for decisions. Independent audits remain useful prior to
downstream application.''
}

\switchcolumn

{\scriptsize
``Post-training \textbf{calibration} aims to align the
\textbf{confidence} produced by a classifier with the observed frequency of
correct predictions in practice. A common calibration correction learns one
\textbf{temperature} from labeled validation examples and rescales the
\textbf{logits} before turning the adjusted results into
\textbf{probabilities}. Because multiplication by the same positive factor
preserves rank, the predicted category remains identical even though
\textbf{overconfidence} may decrease. Only the calibration coefficient is
fitted; the underlying \textbf{predictor} is left untouched. Performance
can nevertheless degrade when the deployment \textbf{distribution} differs
substantially from the data represented during validation. In that setting,
the relationship between internal evidence and error frequency may shift,
so a single temperature no longer provides an adequate remedy. Flexible
mappings can address categorywise behavior or local variation, but they
consume more supervision and introduce greater \textbf{uncertainty}.
Consequently, confidence estimates should be reevaluated following
important shifts in the population on which the classifier operates. This
reassessment matters most when published estimates guide sensitive
operational choices.''
}

\end{paracol}

\begin{paracol}{2}

{\scriptsize
\textit{\textbf{Idea:}
Temperature scaling calibrates classifier confidence globally but can fail
when deployment distributions differ from validation.}
}

\switchcolumn

{\scriptsize
\textit{\textbf{Idea:}
Global calibration can conceal local or shifted-population failures,
requiring renewed confidence assessment during deployment.}
}

\end{paracol}
\end{document}